\documentclass[superscriptaddress,twocolumn,showpacs,amsmath,amssymb]{revtex4}
\usepackage{graphicx}
\usepackage{xcolor}

\def\be{\begin{equation}}
\def\ee{\end{equation}}
\def\bea{\begin{eqnarray}}
\def\eea{\end{eqnarray}}

\renewcommand{\vec}[1]{\mbox{\boldmath $#1$}}

\begin{document}

\title{
Reaction cross sections of the deformed halo nucleus $^{31}$Ne}

\author{Y. Urata}
\author{K. Hagino}
\affiliation{
Department of Physics,  Tohoku University,  Sendai,  980-8578,  Japan}

\author{H. Sagawa}
\affiliation{
Center for Mathematics and Physics,  University of Aizu,  
Aizu-Wakamatsu,  Fukushima 965-8560, Japan}

%%%%%%%%%%%%%%%%%%%%%%%%%%%%%%%%%%%%%%%%%%%%%%%%%%%%%%%%%%%%%%%
%%%  You may repeat \author \address as often as necessary  %%%
%%%%%%%%%%%%%%%%%%%%%%%%%%%%%%%%%%%%%%%%%%%%%%%%%%%%%%%%%%%%%%%

\begin{abstract}
Using the Glauber theory, 
we calculate reaction cross sections for the deformed
halo nucleus $^{31}$Ne. 
To this end, we assume that the 
$^{31}$Ne nucleus takes the 
$^{30}$Ne + $n$ structure. 
In order to 
take into account the rotational excitation of
the core nucleus $^{30}$Ne, 
we employ the particle-rotor model (PRM). 
We compare the results 
to those in the adiabatic limit
of PRM, that is, the Nilsson model, 
and show that the Nilsson model works reasonably well for 
the reaction cross sections of $^{31}$Ne. 
We also investigate the dependence
of the reaction cross sections on the ground state properties
of $^{31}$Ne, such as the deformation parameter and
the $p$-wave component in the ground state wave function.
\end{abstract}


\maketitle

\section{Introduction}

Interaction cross sections as well as reaction 
cross sections are intimately related to the size of nuclei \cite{T85,O01}. 
Using this property, 
the halo structure has been found in some light neutron-rich nuclei. 
This is a spatially extended density distribution of valence neutrons, 
and has been 
first recognized in $^{11}$Li by Tanihata {\it et al.} \cite{T85}.
The root-mean-square radius diverges for $s$ and $p$ waves
as the single-particle energy approaches to zero 
\cite{RJM92}, and the halo structure has been ascribed to an 
occupation of an $l=0$ or $l=1$ orbit by the valence neutron \cite{S92}.
$^{11}$Be \cite{T88,F91} and $^{19}$C \cite{N99} have been 
regarded as $s$-wave halo nuclei,
and $^{6}$He \cite{T85_2} is an example of a $p$-wave halo nucleus.

A large interaction cross section for $^{31}$Ne was recently observed 
by Takechi {\it et al.} \cite{Take12}. 
This observation suggests the extended
density distribution for $^{31}$Ne, that is, the halo structure, 
being 
consistent also with 
a large 
Coulomb breakup cross section measured by Nakamura {\it et al.}\cite{N09}.
Takechi {\it et al.} have analyzed the data 
using single particle levels in a deformed 
potential and 
argued that 
$^{31}$Ne is an $s$- or $p$- wave
halo nucleus\cite{Take12}. 

The ground state properties
for $^{31}$Ne have not been known well. 
For example, the one neutron separation energy
S$_n$ = 0.29 $\pm$ 1.64 MeV \cite{J07} has a large uncertainty and
the spin and parity have not yet been determined. 
In  the core nucleus $^{30}$Ne,  
the candidates 
for the first excited 2$^{+}$
and 4$^{+}$ states have been experimentally observed 
at excitation energies of 0.801 MeV and 2.24 MeV, respectively \cite{D09,F10}.
The energy ratio $E_{4^+}/E_{2^+}$=2.80 suggests
this nucleus to be a transitional one in comparison
with the ratio 3.33 for well-deformed nuclei.
Hamamoto has carried out the Nilsson model calculation with 
a deformed Woods-Saxon potential
and argued that the [330 1/2],
[321 3/2], and [200 1/2] Nilsson levels occupied by the valence
neutron 
can hold 
the halo structure \cite{H10}. 
The [330 1/2] and [321 3/2] configurations 
lead to $I^{\pi}=3/2^-$, while the [200 1/2] leads to 
$I^{\pi}=1/2^+$ for the spin 
and parity of the ground state of $^{31}$Ne in the laboratory frame.

In the previous publication, we used a 
particle-rotor model (PRM) \cite{RS80,BM75,BBH87,EBS95,NTJ96,TH04} 
to analyze the experimental data for the Coulomb breakup cross section 
and discussed the ground state configuration for the $^{31}$Ne 
nucleus \cite{U11}. 
Notice that the Nilsson model corresponds to the adiabatic limit of PRM. 
We have shown that 
the ground state configuration 
corresponding to the [321 3/2] Nilsson orbit 
can be excluded 
if the finite rotational excitation energy of the core nucleus 
is taken into account \cite{U11}. 

In this paper, we apply the same model to the reaction 
cross section of $^{31}$Ne. 
The effect of deformation on 
the reaction cross section of 
the $^{31}$Ne nucleus has been discussed recently 
by Minomo {\it et al.} 
using the 
microscopic optical potential model \cite{M11,M12}, and 
has been shown to play an important role. 
It would thus be of interest to discuss the role of deformation 
in the reaction cross section of $^{31}$Ne using the PRM as an alternative 
approach, 
which has been successful 
in reproducing the Coulomb breakup cross section. 
Notice that 
reaction cross sections with deformed projectiles have been 
evaluated by Christley and Tostevin with the optical limit Glauber 
theory \cite{C99}. We will 
extend it to a system of deformed core nucleus 
plus a valence neutron, based on the 
formalism given in Ref. \cite{B05} for single-nucleon knockout 
reactions. 

The paper is organized as follows. In Sec. II, 
we summarize 
the framework of PRM 
and the calculation procedure for the
reaction cross section. 
In Sec. III, we present the results of the reaction cross section
for $^{30,31}$Ne. 
We discuss the effect of the finite
rotational excitation energy on the reaction cross section.
We investigate also the 
dependence of the reaction cross section 
on the 
deformation, 
the ground state configuration, and the rms radius of $^{31}$Ne.
In Sec. IV, we summarize the paper. 

\section{Formalism}
\subsection{Particle-rotor Model}

\begin{figure}\label{figure_PRM}
\includegraphics[scale=0.5,clip]{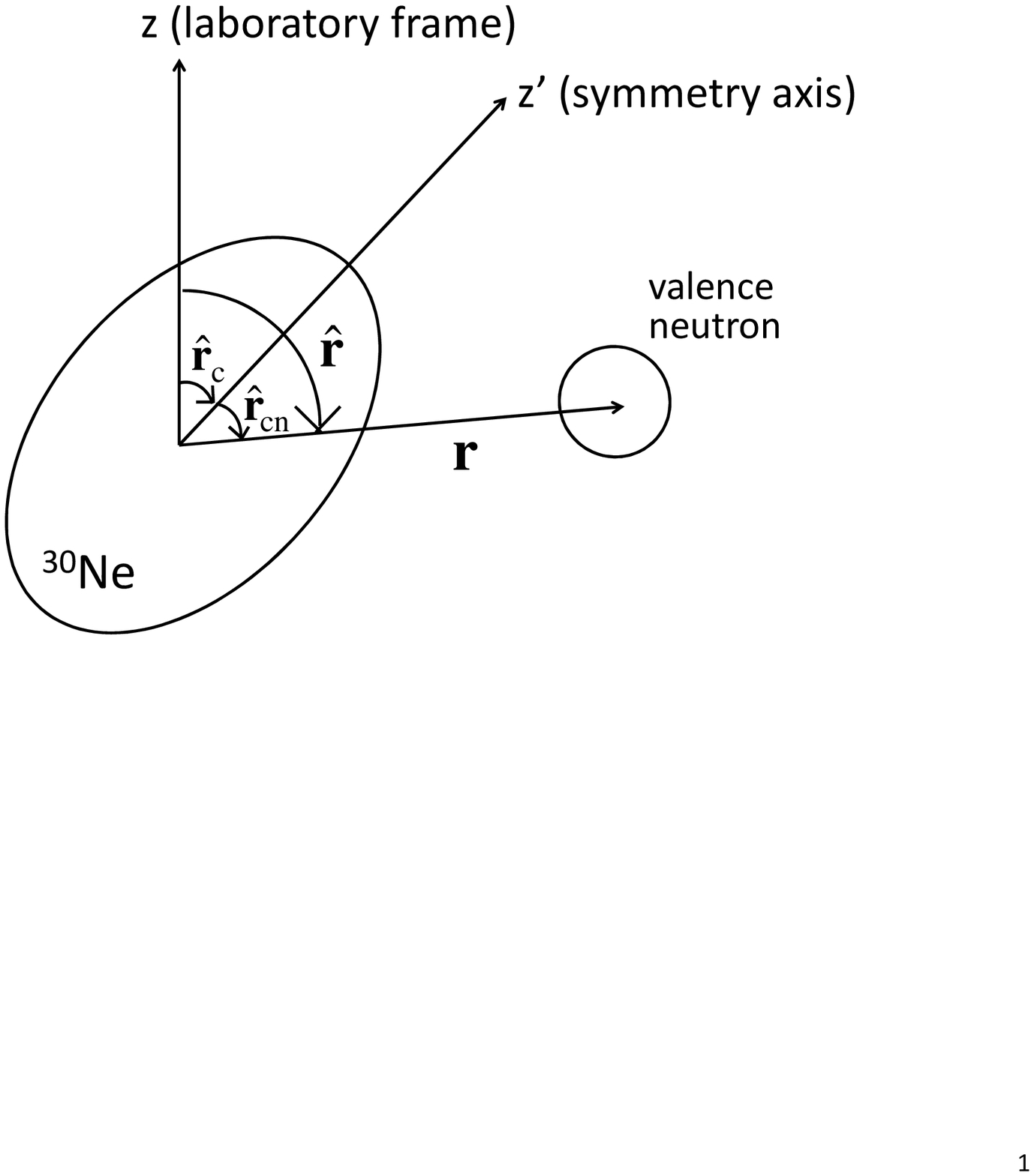}
\caption{(Color online)
The coordinates which define a system of the deformed core 
nucleus $^{30}$Ne plus a neutron. 
$z'$ denotes the symmetry axis for the deformed core nucleus $^{30}$Ne.
$\vec{r}$ and $\hat{\vec{r}}_c$ are the
coordinates for the valence neutron and the direction of the symmetry
axis of the core nucleus in the laboratory frame, respectively.
$\hat{\vec{r}}_{cn}$
is the angle between $\vec{r}$ and $\hat{\vec{r}}_c$.
}
\end{figure}

In order to compute the reaction cross section of the 
$^{31}$Ne nucleus, we assume that it consists of 
the statically
deformed core nucleus $^{30}$Ne and one 
valence neutron as shown in Figure 1. The relevant coordinate 
systems are also shown in the figure.
In this model, 
the single particle motion of the valence neutron is coupled
to the rotation of the deformed core nucleus.
For simplicity, 
we assume the axially symmetric deformation for the core nucleus
with the quadrupole deformation parameter $\beta_{2}$. 
We consider the same Hamiltonian for this system as in Ref. \cite{U11},
\begin{equation}
H = -\frac{\hbar^2}{2\mu}\vec{\nabla}^2+V(\vec{r},\hat{\vec{r}}_c)+H_{\rm rot},
\label{Hprot}
\end{equation}
where $\mu=m_{\rm N}A_c/(A_c+1)$ is the reduced mass of the valence neutron,  
with $A_c$=30 and $m_{\rm N}$ being the mass number of the core nucleus 
and the nucleon mass, respectively. 
$H_{\rm rot}$ is the rotational Hamiltonian for the core nucleus. 
$V(\vec{r},\hat{\vec{r}}_c)$ is the single-particle potential for the valence neutron interacting 
with the deformed core. $\vec{r}$ and $\hat{\vec{r}}_c$ are the
coordinates of the valence neutron and the direction of the symmetry
axis of the core nucleus in the laboratory frame, respectively 
(see Fig. 1). 
We use a deformed Woods-Saxon potential for $V$ and expand it up to the
linear order of the deformation parameter $\beta_{2}$ as,
\be
V(\vec{r},\hat{\vec{r}}_c)\sim V_{0}(r)+V_{\rm def}(r,\hat{\vec{r}}_{cn}), 
\ee
where $V_{0}$ is a spherical Woods-Saxon potential together with 
the spin-orbit ($ls$) 
force, and $V_{\rm def}$ is the deformed part of the potential given by, 
\be
V_{\rm def}(r,\hat{\vec{r}}_{cn})
=-R_0\beta_2\frac{dV_0^{(0)}(r)}{dr}\,Y_{20}(\hat{\vec{r}}_{\rm cn}),
\ee
where $V_0^{(0)}(r)$ is the central part of the 
spherical Woods-Saxon potential, $V_0(r)$, 
and $\hat{\vec{r}}_{cn}$
is the angle between $\vec{r}$ and $\hat{\vec{r}}_c$. 
The deformation of the $ls$ potential is neglected for simplicity. 
We have checked 
the validity of the expansion up to the linear order of $\beta_{2}$ 
by comparing
to the calculation with the higher order terms, 
and have confirmed that it works well. 

Since the calculation of the reaction cross section needs only the ground state
wave function of $^{31}$Ne, it is sufficient
to expand the wave function on the basis 
{\it e.g.,} 
the eigen-functions of
the spherical part $V_{0}$ of the potential,
$R_{njl}(r){\cal Y}_{jlm}(\hat{\vec{r}})$, where
$R_{njl}(r)$ is the radial wave function and ${\cal Y}_{jlm}(\hat{\vec{r}})$
is the spin-angular wave function. 
The continuum spectrum can be discretized within a large box. 
Together with the rotational wave function $\phi_{I_cM_c}(\hat{\vec{r}}_c)$, 
the total wave function for the $n$+$^{30}$Ne system is expanded as, 
\begin{equation}
\Psi_{IM}(\vec{r},\hat{\vec{r}}_c)=\sum_{njl}\sum_{I_c}\alpha^{(I)}_{njlI_c}
R_{njl}(r)[{\cal Y}_{jl}(\hat{\vec{r}})\phi_{I_c}(\hat{\vec{r}}_c)]^{(IM)}, 
\label{wf}
\end{equation}
where $I$ is the spin of $^{31}$Ne and $M$ is its $z$-component. 
The expansion coefficients $\alpha^{(I)}_{njlI_c}$ as well as the 
corresponding eigen-energies for the $^{31}$Ne nucleus are obtained by 
numerically diagonalizing the Hamiltonian $H$. 

We identify the ground state configuration in the same manner as
in the previous work \cite{U11}.
That is, 
we first solve the Hamiltonian in the adiabatic limit 
by setting $H_{\rm rot}=0$ in Eq. (\ref{Hprot}),
that is, by assuming that all the members of the ground rotational 
band are degenerate in energy. 
In this case, the $K$ quantum number, that is, 
the projection of the total 
angular momentum onto the $z$-axis in the body-fixed frame, is conserved, 
and several states with different $I$,
having the same value of $K$, are degenerate in energy when 
the maximum value of $I_c$ included in the calculation is sufficiently large.
The wave function in this limit is related to the wave function 
in the Nilsson model, $\phi_{jlK}$, in the sense that
it is a transformation of the Nilsson 
wave function from the body-fixed frame to 
the laboratory frame. 
The eigen-energies so obtained thus 
form the
single-particle Nilsson levels.
The probability for the ($j,l$) component in the Nilsson wave function is
represented in the following relation:
\begin{equation}
\sum_{I_c}\sum_n |\alpha^{(I)}_{njlI_c}|^2=\int r^2dr \phi_{jlK}(r)^2=P_{jl}^{\rm(Nil)}, 
\label{Pnil}
\end{equation}
which is independent of $I$. 

In order to construct 
the ground state, 
we put two neutrons to each Nilsson orbit from the 
bottom of the potential well, and seek the Nilsson orbit 
which is occupied by the last 
unpaired neutron. We then gradually 
increase the value of the 2$^+$ energy of the core nucleus up to the 
physical value, $E_{2^+}$=0.801 MeV, and 
monitor how the Nilsson orbit for the valence neutron evolves. 
For a finite value of $E_{2^+}$, the $K$ quantum number is not conserved 
any more due to the Coriolis coupling, and 
the degeneracy with respect to $I$ is resolved. 
We select the lowest energy state among several $I$ at 
$E_{2^+}$=0.801 MeV as the ground state of $^{31}$Ne. 
In this way, we take into account the Pauli principle between 
the valence neutron and the neutrons in the 
core nucleus.

We 
consider 
two configurations with the spin-parity of $I^{\pi}=3/2^{-}$ in
the laboratory frame as candidates for the ground state of $^{31}$Ne.
One with the deformation parameter $\beta_{2}=0.2$
corresponds to the Nilsson level [330 1/2],
and the other with $\beta_{2}=0.55$ corresponds to the Nilsson
level [321 3/2] in the adiabatic limit where the rotational
energy of the core nucleus is neglected.
Following Ref. \cite{M11}, 
we use the same Woods-Saxon potential parameters as 
those given in Table I of Ref. \cite{S09}. 
The depth of the Woods-Saxon potential is varied 
to reproduce the one neutron separation energy. 
We use a similar value for the energy cut-off 
for the single particle basis 
and a similar size of the box to discretize the
continuum spectrum as in Ref. \cite{U11}.

\subsection{Reaction Cross Sections}

In this paper, we discuss 
the reaction cross sections for $^{30,31}$Ne on the
carbon target.
For simplicity, we neglect 
the effect of the Coulomb force. 
To verify this approximation, we have calculated the 
Coulomb breakup cross sections
of $^{31}$Ne for the configuration with $\beta_2$=0.2
at the separation energy $S_n$=0.2MeV on the carbon target 
using the method given by Ref. \cite{U11}.
The calculated Coulomb breakup cross section, 0.0033b,
is indeed small as compared to the total reaction cross section, 
suggesting that the nuclear force dominantly contributes
to reaction cross sections for $^{31}$Ne on the carbon target. 
In order to compute the reaction cross sections, 
we use the Glauber theory where the eikonal approximation
and the adiabatic approximation are adopted \cite{G59}.
We closely follow the formalism in Ref. \cite{B05}, in which 
the PRM has been used to evaluate 
single-nucleon knockout reactions of a deformed odd-A nucleus 
based on the Glauber theory. 

In the eikonal approximation, the final state $\Psi_{f}$
after the collision 
is described with the initial state wave function $\Psi_{i}$ as,
\be
\Psi_{f}=S\ \Psi_{i}=\exp[i\chi ]\ \Psi_{i},
\ee
where $\chi$ is the phase shift function. 
The reaction cross section of $^{31}$Ne, defined 
as the difference between the total cross section and 
the elastic scattering cross section, 
is given
with the ground state wave function of $^{31}$Ne,
$\Psi_{IM}$, as,
\bea
&&\sigma_{R}(^{31}{\rm Ne})=\int d\vec{b}\,\biggl(1-\frac{1}{2I+1}\sum_{M} \notag \\
&&\hspace{85pt}\times\bigl|\ \langle\Psi_{IM}\left|
\,S_{c}S_{v}\,\right|\Psi_{IM}\rangle\,\bigr|^{2}\biggr) ,\ \  \label{sgm31}
\eea
where $\vec{b}$ is the impact parameter
of the center of mass of the projectile nucleus $^{31}$Ne 
colliding with the target nucleus.
The $S$-matrix for the two-body projectile
nucleus can be written by $S\sim S_{c}S_{v}$ in the 
Glauber approximation \cite{AKTT96,HBE96}.
Here, 
$S_{c}$ and $S_{v}$ are $S$ matrices for the core nucleus and the valence neutron,
respectively. 
Notice that, since the directions of $\vec r$ and $\hat{\vec{r}}_{c}$ 
are integrated in the whole space 
in Eq. (\ref{sgm31}) before the integration over $\vec{b}$ is 
carried out, the integrand does not depend upon the direction 
of $\vec{b}$ \cite{B05}. 
The reaction cross section $\sigma_{R}(^{31}{\rm Ne})$ thus reads
\bea
&&\sigma_{R}(^{31}{\rm Ne})=2\pi \int b\,db\,\biggl(1-\frac{1}{2I+1}\sum_{M} \notag \\
&&\hspace{50pt}\times\biggl|\int d\vec{r}\int d\hat{\vec{r}}_{c}
\,S_{v}S_{c}\,F(\vec{r},\hat{\vec{r}}_c)\biggr|^{2}\,\biggr), \label{sgm31_1}
\eea
where 
\bea
F(\vec{r},\hat{\vec{r}}_c)
&=&\sum _{n'j'l'I_{c}'}\sum _{njlI_{c}}\alpha ^{*}_{n'j'l'I_{c}'}\alpha _{njlI_{c}}
R^{*}_{n'j'l'}(r)R_{njl}(r) \notag \\
&\times &
\sum_{m_{j}'m_{I}'m_{j}m_{I}}
\langle j'm_{j}'I_{c}'m_{I}'|IM\rangle \langle jm_{j}I_{c}m_{I}|IM\rangle \notag \\
&\times &{\cal Y}_{j'l'm_j'}^{*}(\hat{\vec{r}})\phi_{I_c'}^{*}(\hat{\vec{r}}_c)
{\cal Y}_{jlm_j}(\hat{\vec{r}})\phi_{I_c}(\hat{\vec{r}}_c).
\eea
Using the formula for the product of two spherical harmonics
with the same angles,
the function $F$ is transformed to \cite{B05}
\be
F(\vec{r},\hat{\vec{r}}_c)
=\sum_{L,m_L,{\cal L},m_{\cal L}}
Y_{Lm_L}(\hat{\vec{r}})Y_{{\cal L}m_{\cal L}}(\hat{\vec{r}}_c)U_{Lm_L{\cal L}m_{\cal L}}(r),
\ee
where
\bea
U_{Lm_L{\cal L}m_{\cal L}}(r)
&=&\sum _{n'j'l'I_{c}'}\sum _{njlI_{c}}\,\sum_{km_k}
\alpha ^{*}_{n'j'l'I_{c}'}\,\alpha _{njlI_{c}} \notag \\
&\times &R^{*}_{n'j'l'}(r)R_{njl}(r) \notag \\
&\times &(-)^{2j'-j+I_c'+m_k-\frac{1}{2}}
\frac{\hat{I}\,\hat{j}'\,\hat{j}\,\hat{l}'\,\hat{l}\,\hat{I_{c}}'\hat{I_{c}}\,\hat{k}}{4\pi} \notag \\
&\times &\langle l\,0\,l'\,0\,|L\,0\rangle \langle I_{c}0\,I_{c}'0|{\cal L}\,0\rangle \notag \\
&\times &\langle IM\,k-m_k|IM\rangle \langle Lm_L{\cal L}m_{\cal L}|km_k\rangle \notag \\
&\times &W(\,l'\,j'\,l\,j\,;\frac{1}{2}L)
\begin{Bmatrix}I_c' &I_c &{\cal L} \\
               j'   &j   &L        \\
               I    &I   &k        
\end{Bmatrix}, 
\eea
with $\hat{j}=\sqrt{2j+1}$ and $W$ being the Racah coefficients. 

In order to evaluate the $S$-matrices, we employ the optical limit
approximation for simplicity.
Using the zero range
interaction for the effective nucleon-nucleon interaction, we evaluate
the $S$ matrices 
by folding the densities of the projectile and the target nuclei as,
\bea
&S_{c}&=\exp[-\bar\sigma_{\rm{NN}}(1-i\bar\alpha_{\rm{NN}})
\chi_{c}(\vec{b}_c,\hat{\vec{r}}_c)/2], \label{sc} \\
&S_{v}&=\exp[-\bar\sigma_{\rm{NN}}(1-i\bar\alpha_{\rm{NN}})\chi_{n}(b_n)/2], \label{sn}
\eea
where
\bea
&\chi_{c}(\vec{b}_c,\hat{\vec{r}}_c)
&=\int dz_{c}\int d\vec{r}'\rho_{c}(\vec{r}'
,\hat{\vec{r}}_{c})
\rho_{T}(|\vec{r}'+\vec{R_{c}}|), \label{chic} \\
&\chi_{n}(b_n)&=\int dz_{n}\rho_{T}(R_{n}). \label{chin}
\eea
Here, 
$\vec{R}_{c}=(\vec{b}_c,z_c)$ 
and $\vec{R}_{n}=(\vec{b}_n,z_n)$ 
are the coordinates of the center of mass 
of the core nucleus and the valence neutron 
from the target nucleus, respectively. 
$\rho_{c}$ and $\rho_{n}$ are the densities of the core and the 
target nuclei, respectively.
We construct the density of the core nucleus $^{30}$Ne 
with the Nilsson model. 
To this end, 
we use the original values for the 
potential parameters
given in Table I of Ref. \cite{S09}. 
For the density distribution for the target nucleus
 $^{12}$C, 
we use a one-range Gaussian function whose width parameter 
is determined so as to reproduce 
the experimental root mean square radius. 
$\bar\sigma_{\rm{NN}}$ 
in Eqs. (\ref{sc}) and (\ref{sn}) 
is the average value 
of the total cross sections of the 
nucleon-nucleon scattering \cite{R79}. 
$\bar\alpha_{\rm{NN}}$ is also the average value of the ratio of the real to
the imaginary part of the nucleon-nucleon scattering amplitudes.
We use the experimental values for $\sigma_{pp}$, $\sigma_{pn}$, $\alpha_{pp}$ 
and $\alpha_{pn}$ for the incident energy 240 MeV/nucleon 
listed in Ref. \cite{A08}.

Notice that 
the reaction cross section of the core nucleus $^{30}$Ne is simply given by 
\be
\sigma_{R}(^{30}{\rm Ne})=\int d\vec{b_{c}}\frac{1}{4\pi}\int d\vec{r_{c}}
\left( 1-\left| S_{c}\right| ^{2}\right), \label{sgm30}
\ee
with $S_{c}$ given in Eqs. (\ref{sc}) and (\ref{chic}) \cite{C99}.

\section{Results}

\begin{figure}\label{conf_exp}
\includegraphics[scale=0.55,clip]{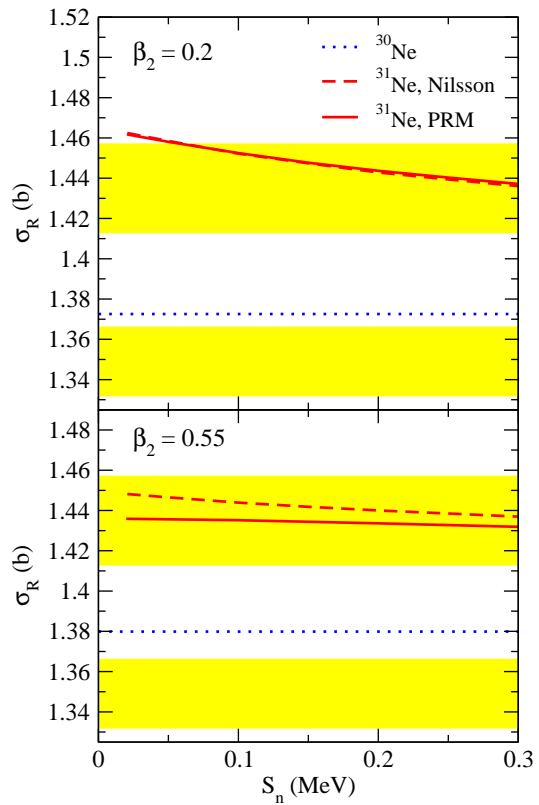}
\caption{(Color online)
The reaction cross sections of $^{30,31}$Ne 
obtained with the particle-rotor model 
as a function of the
one neutron separation energy for $^{31}$Ne, $S_{n}$.
The upper and the lower panels show the results for 
the configuration with $\beta_{2}$=0.2 and 0.55
for $^{31}$Ne, respectively. The dashed lines are the calculations 
in the adiabatic limit, while
the solid lines are the calculations with the finite rotational energy.
The dotted lines denote the reaction cross sections for $^{30}$Ne.
The upper and the lower shaded regions indicate the experimental 
interaction cross sections for 
$^{31}$Ne and $^{30}$Ne, 
respectively.
}
\end{figure}

\begin{figure}
\includegraphics[scale=0.55,clip]{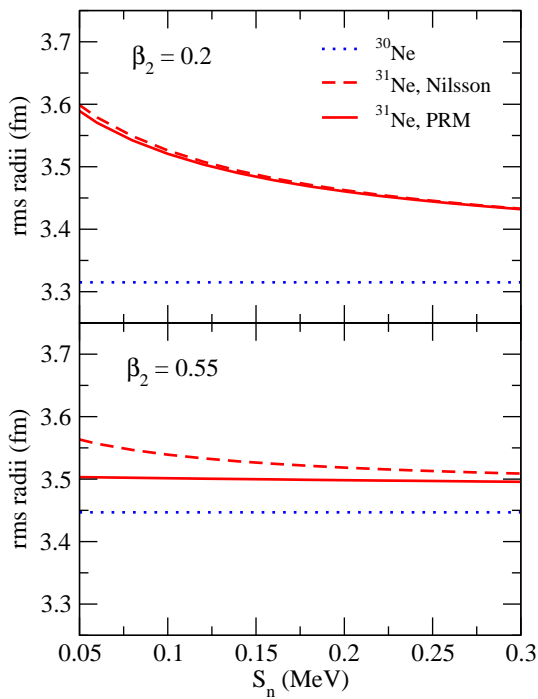}
\caption{(Color online)
The root mean square radii of $^{30,31}$Ne as a function of the
one neutron separation energy. 
The meaning of each line is the same 
as in Fig. 2.
}
\end{figure}

We now numerically evaluate the reaction cross sections 
for $^{30,31}$Ne. 
The upper and the lower panels of Fig. 2 show the results for 
the configurations
with $\beta_{2}=0.2$ and 0.55, respectively. 
Since the measured one-neutron separation energy $S_{n}$
of $^{31}$Ne has a large error bar, 
S$_n$ = 0.29 $\pm$ 1.64 MeV \cite{J07}, 
we show the calculated 
reaction cross sections as a function
of $S_{n}$. 
The upper and the lower shaded regions in each panel 
indicate the experimental interaction cross sections 
for $^{31}$Ne and $^{30}$Ne\cite{Take12}, respectively.
The dashed lines are the calculations in the adiabatic limit, while
the solid lines take into account the finite rotational energy of the 
core nucleus.
We show also the reaction cross sections for $^{30}$Ne with the dotted lines.

As one can see, the results of PRM are similar to those in the adiabatic 
limit 
for both the configurations with $\beta_{2}$=0.2 and 0.55. 
In the adiabatic limit, since
each component of ($j,l,I_{c}$) with different values of $I_{c}$ has the same 
radial wave function, it is the 
total $p_{3/2}$ probability, summed with different $I_{c}$ values, that 
is relevant to the halo structure. 
On the other hand, due to the non-adiabatic effect, the wave function
[$I_{c}=0^{+}\otimes p_{3/2}$] 
is spatially most extended in the PRM\cite{U11}. 
For the configuration with $\beta_{2}=0.2$ and $S_{n}=0.2$ MeV,
the probability for the total $p_{3/2}$ component
is 54.9$\%$ in the adiabatic limit, which is 
almost equal to the probability for the [$0^{+}\otimes p_{3/2}$] 
component in the PRM, that is, 54.2$\%$. 
The halo structure therefore retains even when the finite 
excitation energy is taken into account in the PRM. 
For the configuration with $\beta_{2}=0.55$, on the other hand, 
the total $p_{3/2}$ probability in the adiabatic limit 
is 25.7\% 
and the probability for the 
[$0^{+}\otimes p_{3/2}$] component 
in the PRM is 2.1 $\%$. 
Therefore, the halo structure disappears for this configuration when 
the finite rotational energy is taken into account\cite{U11}. 
The small difference between the solid and the dashed curves in the lower 
panel of Fig.2 reflects this fact. 
Nevertheless, the halo contribution to the reaction cross section 
does not seem large in this mass region, 
and the adiabatic approximation still works for 
the reaction cross sections. 

In order to see the relation between the halo structure and the 
reaction cross section more clearly, Fig. 3 shows 
the rms radii for those configurations 
as a function of $S_{n}$. 
The behaviors of the rms radii are qualitatively the same as
the reaction cross sections shown in Fig. 2. 
The rms radius is almost constant for $\beta_2$=0.55 when the finite 
rotational energy is taken into account, that is consistent with 
the disappearance of the halo structure. The rms radii increase 
as the one-neutron separation energy, $S_n$, decreases for the other 
cases, indicating the halo structure. 
Notice that in contrast to the rms radii and the Coulomb breakup 
cross section, 
the reaction cross section is less sensitive 
to the extended
density distribution, since the inner part of the density distribution 
also contributes to the cross section. 

The calculated density distribution of $^{31}$Ne
for the configuration with $\beta_2$=0.2 
and $S_n$=0.2 MeV
is shown in Figure 4.
The dashed and the solid lines are the results for $^{30}$Ne and $^{31}$Ne, respectively.
Since the effects of the finite excitation
of the core nucleus on the reaction cross sections and rms radii are 
small for this configuration,
we calculate the density for $^{31}$Ne in the adiabatic limit, that is, Nilsson model,
which is defined in the body-fixed frame.
The upper and the lower panels show the density distributions in the direction
of the symmetry axis of the core nucleus $^{30}$Ne and in the direction perpendicular
to the symmetry axis, respectively.
The density distribution 
has
an exponentially extended tail, 
indicating the halo structure for this nucleus. 
Notice that the density distribution is proportional to 
$Y_{00}(\theta_{\rm cn})+Y_{20}(\theta_{\rm cn})/\sqrt{5}$ 
for the pure $p_{3/2}$ 
state with $K=1/2$, and it is extended more 
in the direction of the 
symmetry axis compared to the direction perpendicular to it. 

\begin{figure}%[!b]
\includegraphics[scale=0.55,clip]{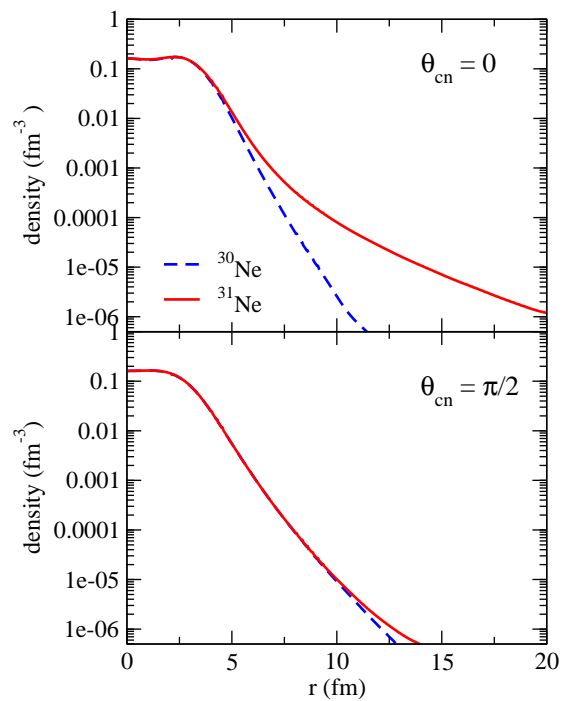}
\caption{(Color online)
The density distribution for $^{31}$Ne
with $\beta_2$=0.2 and $S_n$=0.2 MeV obtained with 
the adiabatic limit of the particle-rotor model, that is, the 
Nilsson model.
$\hat{\vec{r}}_{\rm cn}=(\theta_{\rm cn},\phi_{\rm cn})$ 
is the angle from the symmetry axis of the core nucleus.
The upper and the lower panels show the results in the direction of
the symmetry axis of the core nucleus $^{30}$Ne and in the direction perpendicular
to the symmetry axis, respectively.
}
\end{figure}

In Figure 2, the calculated reaction cross sections appear to 
reproduce 
the experimental data
for both the configurations with $\beta_2$=0.2 and 0.55. 
However, the increase of the calculated
reaction cross section from
$^{30}$Ne to $^{31}$Ne 
is much smaller for the configuration with 
$\beta_{2}=0.55$ than with $\beta_{2}=0.2$.
This is because the probability of the $p_{3/2}$ component
is much larger at $\beta_{2}=0.2$ than at $\beta_{2}=0.55$.
Notice that these  different  $p_{3/2}$ probabilities 
stem  from the  Nilsson levels 
[330 1/2] at $\beta_{2}=0.2$ and [321 3/2] at $\beta_{2}=0.55$
in the adiabatic limit, respectively.
The difference between the reaction cross sections
for $^{31}$Ne and $^{30}$Ne 
may be approximately identified as 
the one-neutron
removal cross section for $^{31}$Ne\cite{HSCB10},
\be
\sigma_{-1n}(^{31}{\rm Ne})\sim\sigma_{R}(^{31}{\rm Ne})-\sigma_{R}(^{30}{\rm Ne}). 
\label{rem}
\ee
Figure 5 shows the one-neutron removal cross sections for $^{31}$Ne
on the carbon target so obtained as a function
of the one-neutron separation energy for $^{31}$Ne.
The shaded region indicates the experimental data
with the incident energy of 230 MeV/nucleon \cite{N09}.
The thick and the thin lines are the one-neutron removal cross sections
for the configuration with
$\beta_2$=0.2 and 0.55, respectively. 
The dashed and the solid lines are
the results in the adiabatic limit and with the finite rotational excitation,
respectively.
One can clearly see that 
the results with the configuration with $\beta_2$=0.2 reproduces 
the experimental data, while the configuration with 
$\beta_2$=0.55 is inconsistent with the experimental one-neutron 
removal cross section. 
We conclude that the configuration with $\beta_{2}=0.2$ is 
a very promising candidate for 
the ground state of the deformed halo nucleus $^{31}$Ne, 
which is consistent
with the analysis of the Coulomb dissociation cross section 
of $^{31}$Ne with the PRM \cite{U11}.

\begin{figure}
\includegraphics[scale=0.55,clip]{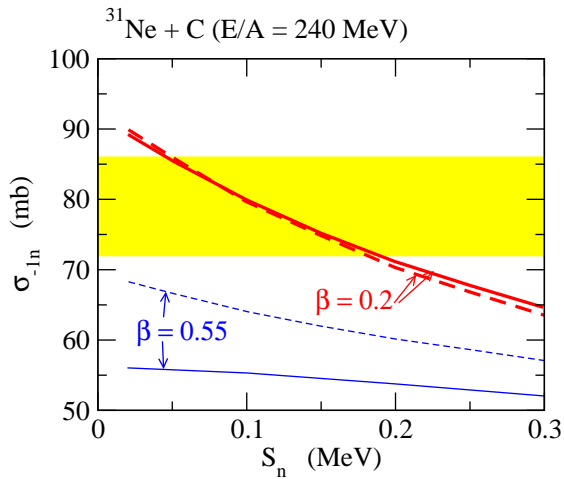}
\caption{(Color online)
The one-neutron removal cross sections of $^{31}$Ne 
defined as a difference between the reaction cross sections
for $^{31}$Ne and $^{30}$Ne 
as a function of the
one neutron separation energy for $^{31}$Ne, $S_{n}$.
The thick and thin lines are the results for the configuration
with $\beta_2$=0.2 and 0.55 for $^{31}$Ne, respectively.
The dashed lines are the calculations 
in the adiabatic limit
of the particle-rotor model, while
the solid lines are the calculations with the finite rotational energy.
The shaded region indicates the experimental 
one-neutron removal cross section for $^{31}$Ne, taken from 
Ref. \cite{N09}.
}
\end{figure}

\begin{figure}
\includegraphics[scale=0.55,clip]{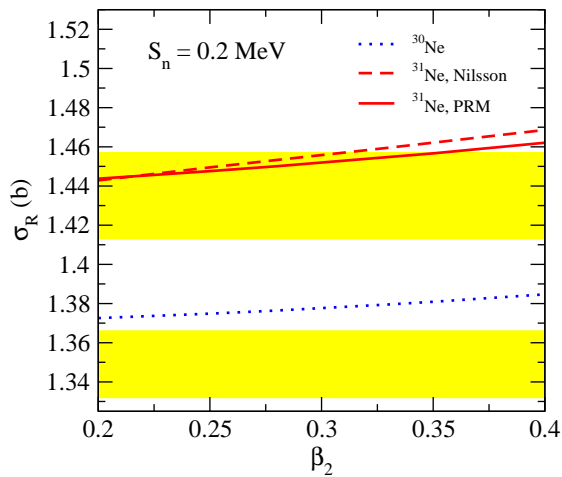}
\caption{(Color online)
The reaction cross sections for $^{30,31}$Ne 
as a function of the deformation parameter $\beta_{2}$.
For $^{31}$Ne, 
the one neutron
separation energy is assumed to be $S_{n}=0.2$ MeV, with 
the configuration 
corresponding to that in the upper panel of Fig. 2. 
The meaning of each line is the same as in Fig. 2. }
\end{figure}

We next investigate the deformation dependence of the reaction
cross sections of $^{30,31}$Ne 
for the configuration which reproduces the experimental data
of the reaction and the one-neutron removal cross section at $\beta_{2}=0.2$. 
With the method explained in section IIA,
this configuration remains the ground state 
in the range of 
the deformation parameter, 0.17$\lesssim\beta_2\lesssim$0.33.
Given the uncertainties of the potential parameters,
this configuration 
may be the ground state even at around $\beta_{2}=0.4$, 
as suggested by the Anti-symmetrized Molecular Dynamics (AMD) 
calculation for $^{29-31}$Ne
\cite{K04,K07}, see Table II in Ref. \cite{M11}.
Figure 6 shows the reaction cross sections for $^{30,31}$Ne
for the separation energy of $S_{n}=0.2$ MeV, as a function of 
the deformation parameter in the region
of $0.2\leq\beta_{2}\leq0.4$.
The dashed line is the result in the adiabatic limit, while 
the solid line is the result with the finite rotational energy.
The result for $^{30}$Ne is shown with the dotted line.
The reaction cross sections for $^{30,31}$Ne smoothly increase
only by about 0.01 b 
from $\beta_2=0.2$ to $\beta_2=0.4$
due to the deformation of the core density. 
This may be understood in terms of the deformation dependence 
of the rms radius, see {\it e.g.,} Eq. (1) in Ref. \cite{C99}. 
The total $p_{3/2}$ probability in the adiabatic limit varies from 
54.9\% to 57.0\% as the deformation parameter changes from 
$\beta_2=0.2$ to $\beta_2=0.4$. 
Consequently, 
the deformation dependence of the reaction cross section
of $^{31}$Ne is small, as far as the same configuration is concerned
,i.e., the Nilsson level [330 1/2] in the adiabatic limit. 
The experimental reaction cross section
can thus be reproduced 
within the region of $0.2\lesssim\beta_{2}\lesssim 0.4$.

\section{Summary}

We have discussed the reaction cross sections
for $^{30,31}$Ne 
with the particle-rotor model.
Assuming the system with the deformed core nucleus and 
one valence neutron for $^{31}$Ne, 
the finite rotational excitation
energy of the core nucleus $^{30}$Ne 
is taken into account.
In order to calculate the
reaction cross section on the carbon target, we have used 
the optical limit approximation of the Glauber theory.
We have considered two configurations with the spin-parity of 
$I^{\pi}=3/2^{-}$ at $\beta_{2}=$ 0.2 and 0.55 
as candidates for the ground state of $^{31}$Ne,
 corresponding 
to the Nilsson levels [330 1/2] and [321 3/2] in the adiabatic limit,
  respectively.
The effect of the finite rotational energy changes
the probability of each component in the wave function, especially the
proportion of the [$0^{+}\otimes p_{3/2}$] component as well as the
[$2^{+}\otimes p_{3/2}$] component.
We have found that the non-adiabatic effects on the reaction cross sections
for these two configurations are small, 
and it is concluded
that the Nilsson model works reasonably well for the
reaction cross section for $^{31}$Ne.
We have also found that 
the difference of the reaction cross sections between $^{31}$Ne and 
$^{30}$Ne is much larger for 
the configuration with $\beta_{2}=0.2$ than for 
the configuration with $\beta_{2}=0.55$, 
leading to a consistent description for one-neutron 
removal cross section for $\beta_{2}=0.2$.

Interaction cross sections of Ne isotopes have been measured 
from $^{20}$Ne to $^{32}$Ne by Takechi {\it et al.}\cite{Take12}. 
The data show 
a large odd-even staggering for $^{30,31,32}$Ne, which has been understood
in terms of the pairing anti-halo effect \cite{H11}.
As we have found, the adiabatic approximation 
works well for the reaction cross sections for neutron-rich Ne isotopes. 
It would thus be interesting 
to describe the deformed nucleus $^{32}$Ne
with {\it e.g., } the Hartree-Fock-Bogoliubov (HFB) method 
taking into account the
pairing interaction and then evaluate the interaction cross section 
in the adiabatic approximation of 
the Glauber theory. 
A work toward this direction is now in progress. 

\subsection*{Acknowledgments}

This work was supported by the Global COE Program
``Weaving Science Web beyond Particle-Matter Hierarchy'' at
Tohoku University,
and by the Japanese
Ministry of Education, Culture, Sports, Science and Technology
by Grant-in-Aid for Scientific Research under
the program numbers (C) 22540262 and 20540277.

\end{document}